\documentclass[
    aps,
    prl,
    reprint,
    amsmath,
    amssymb,
    superscriptaddress
]{revtex4-2}

\usepackage{newtxtext,newtxmath}

\usepackage[dvipsnames]{xcolor}
\usepackage{graphicx}
\graphicspath{{fig/}}
\usepackage[
    colorlinks,
    allcolors=blue
]{hyperref}

\newcommand{\iu}{\mathrm{i}\mkern1mu}
\newcommand{\eu}{\mathrm{e}\mkern1mu}
\renewcommand{\Re}{\operatorname{Re}}

\newcommand{\ort}[1]{\mathbf{\hat{#1}}}
\newcommand{\CDmode}{\mathrm{CD}_{\text{mode}}}
\newcommand{\CDtot}{\mathrm{CD}_{\text{tot}}}
\newcommand{\stp}{\sigma^{\text{top}}}
\newcommand{\sbt}{\sigma^{\text{bot}}}
\newcommand{\grad}{\gamma_{\text{rad}}}
\newcommand{\gabs}{\gamma_{\text{abs}}}
\newcommand{\gtot}{\gamma_{\text{tot}}}
\newcommand{\Bone}{B$_1$}
\newcommand{\Btwo}{B$_2$}
\newcommand{\fluence}{\,\text{\textmu J}/\text{cm}^2}

\usepackage[normalem]{ulem} %

\newcommand{\affsdnu}{Shandong Provincial Key Laboratory of Light Field Manipulation Physics and Applications \& School of Physics and Optoelectronics, Shandong Normal University, Jinan 250358, China}
\newcommand{\affanu}{Research School of Physics, Australian National University, Canberra, ACT 2601, Australia}

\begin{document}

\title{Observation of Chiral Lasing from Monoclinic Resonant Metasurfaces}

\author{Zhuang Li}
\altaffiliation{These authors contributed equally to this work}
\affiliation{\affsdnu}

\author{Ivan Toftul}
\email[]{ivan.toftul@anu.edu.au}
\altaffiliation{These authors contributed equally to this work}
\affiliation{\affanu}

\author{Yangjian Cai}
\email[]{yangjiancai@sdnu.edu.cn}
\affiliation{\affsdnu}

\author{Xianyu Ao}
\email[]{aox@sdnu.edu.cn}
\affiliation{\affsdnu}

\author{Yuri Kivshar}
\affiliation{\affanu}

\date{\today}

\begin{abstract}
We observe chiral lasing from monoclinic lattices of {\it achiral} dielectric pillars placed on a slab. Shearing the lattice breaks its in-plane mirror symmetries and sets the handedness, while preserving twofold rotational symmetry. The lattice dimerization folds guided modes into the light cone and sets their quality factor without changing the handedness. The twofold symmetry keeps a bound state in the continuum (BIC) at the $\Gamma$ point, and its band inherits handedness from the bright mode coupled to it.
We employ this novel approach for chiral metaphotonics to demonstrate experimentally lasing action with two opposite handedness and  large degrees of circular polarization (0.88 and 0.96) originated from BIC band and bright mode, respectively. 
\end{abstract}

\maketitle

A resonant source of circularly polarized light such as a chiral laser relies on a mode that stores light for many optical cycles with a high quality factor ($Q$ factor), and radiates light of one handedness toward a detector~\cite{ScienceChirallasing,adpr.202400060, Gromyko2024, acsnano.5c04290}.
Along any direction that lies in a mirror symmetry plane, left- and right-circularly polarized light is emitted equally, unless the emitters or the excitation break that mirror symmetry~\cite{Light2020, Lai2025, Sun2026}. Sources with achiral emitters therefore need structures without such mirror symmetries.
In metasurfaces, which are planar arrays of subwavelength resonators, symmetry allows cross-polarized circular dichroism once the in-plane mirror symmetries (mirror planes that contain the surface normal) are removed and the rotation symmetry is at most twofold~\cite{Toftul2026arXivCDcross}, whereas co-polarized circular dichroism requires three-dimensional chirality~\cite{PRLYuri2020, PRLAA2021, LightAT2023, lpor.202200611, lpor.202200597, NLYC2024}.
Strong co-polarized circular dichroism with high $Q$ values is often achieved with chiral quasi-bound states in the continuum (BICs). Asymmetry turns a BIC mode into a radiating state splitting its polarization vortex into two points of circular polarization ($C$ points)~\cite{PhysRevLett.123.116104, PhysRevLett.125.053902, Yin2020}, and one $C$ point is then moved to the $\Gamma$ point~\cite{ScienceChirallasing, NatureCQ2023, acsphotonics.2c01926, NClixp2022, NLUVlasing}. The same asymmetry then sets both $Q$ and the handedness, and these two factors are hard to tune independently.
Other approaches suggest to couple two modes through a structural perturbation~\cite{LightCQ2018, PRLChiralFlatband, Deng2025}, twist a photonic-crystal slab~\cite{LightTwistedPCS, LPRQS2026}, merge accidental BICs~\cite{eLightRobustChiral}, or use magneto-optical materials~\cite{PRLMOBIC, SAMOBIC, pnas.2536341123, x4bj-tynz}.
In the lattices with two meta-atoms per the unit cell, band folding sets $Q$ separately from the handedness, which then comes from planar-chiral meta-atoms~\cite{NLSC2025, AMSC2026}.

\begin{figure}[!b]
    \includegraphics[width=2.7432in]{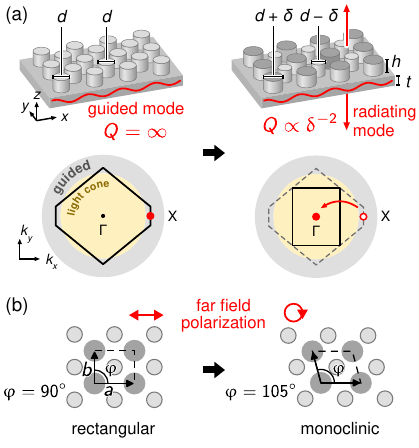}
    \caption{\label{fig:concept}
    Two-step strategy for designing high-$Q$ chiral metasurfaces.
    (a)~Dimerization. 
    With identical pillars of diameter $d$, the lattice is centered, and the structure supports guided modes below light line ($Q = \infty$ for lossless materials).
    For $d\pm\delta$, the pillars double the primitive cell and halve the Brillouin zone. The state at X folds onto $\Gamma$, inside the light cone (yellow), and radiates with $Q\propto\delta^{-2}$. Guided region marked with gray. (b)~Shear. Changing the lattice angle $\varphi$ from rectangular to monoclinic keeps $C_2$ symmetry and changes far-field polarization from linear to elliptical.
    }
\end{figure}

In this Letter, we employ the handedness of the metasurface lattice instead of chirality of individual meta-atoms. We notice that in monoclinic lattices composed of achiral resonators~\cite{Toftul2024PRL, Sinev2025}, dimerization and shear control two different symmetries. Dimerization breaks a translation of the lattice and folds guided modes into the light cone, with $Q\propto\delta^{-2}$, where $\delta$ is the dimerization parameter [see Fig.~\ref{fig:concept}(a)]. Quite differently, shear removes the in-plane mirror symmetries and sets the handedness, while keeping the twofold rotation symmetry $C_2$.  Here we employ these two control mechanisms to demonstrate experimentally chiral metasurface lasing. The ``chiral BIC'' designs mentioned above break the symmetry that protects the BIC~\cite{PRLYuri2020, PRLAA2021, NatureCQ2023} or time reversal~\cite{PRLMOBIC, x4bj-tynz}. Here $C_2$ and time reversal stay, the BIC stays dark, and its band inherits the handedness of the bright modes.

\begin{figure*}[!t]
	\includegraphics[width=\linewidth]{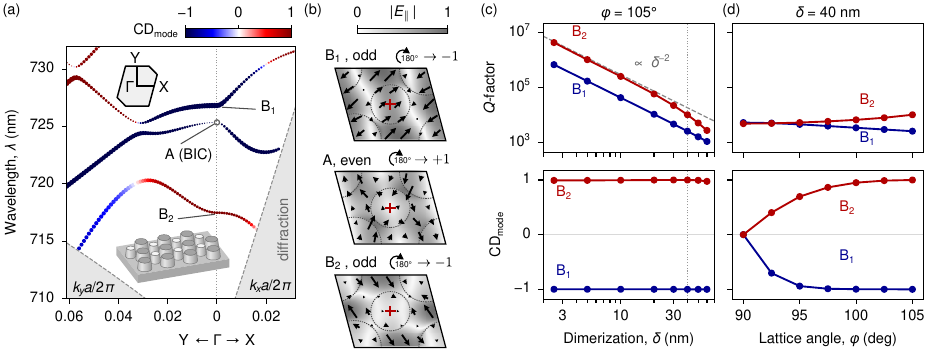}
	\caption{\label{fig:knobs}
        Demonstration of two almost independent control parameters of the radiative $Q$ factor and modal chirality by varying the dimerization $\delta$ and lattice angle $\varphi$ , with the pillar diameter fixed at $d=235$\,nm.
		(a)~Bands near the $\Gamma$ point at $\delta=40$~nm and $\varphi=105^\circ$, colored by $\CDmode$ [Eq.~\eqref{eq:CD}]; the marker area grows with the linewidth $Q^{-1}$. Grey area shows the first diffraction order is open. Open circle: BIC. Insets: Brillouin zone and structure.
		(b)~$C_2$ parity of the three states at the $\Gamma$ point: $|\mathbf{E}_\parallel|$ (grey) and $\Re\mathbf{E}_\parallel$ (arrows) in one cell, on the plane of the pillar bases; $+$: the $C_2$ center.
		(c)~$Q$ (top) and $\CDmode$ (bottom) of \Bone{} and \Btwo{} at the $\Gamma$ point vs. $\delta$, at $\varphi=105^\circ$. Dashed: $Q\propto\delta^{-2}$. Dotted: $\delta=40$~nm, as in (a). (d)~Same but vs. $\varphi$, at $\delta=40$~nm.
	}
\end{figure*}

The metasurface is a slab with cylindrical pillars on its top surface (Fig.~\ref{fig:concept}). The lattice vectors are $\mathbf{a}_1 = a\ort{x}$ and $\mathbf{a}_2 = b(\ort{x}\cos\varphi + \ort{y}\sin\varphi)$. Each cell holds a pillar of diameter $d+\delta$ at its corner and one of diameter $d-\delta$ at its center. 
At $\delta=0$, the centering vector $(\mathbf{a}_1+\mathbf{a}_2)/2$ is a lattice translation, and the slab modes that fold to $\Gamma$ lie below the light line as bound guided modes.
A nonzero $\delta$ breaks this translation, doubles the primitive cell, and folds these modes to $\Gamma$, inside the light cone [Fig.~\ref{fig:concept}(a)]~\cite{zhuPRB2026, acsphotonics.5c00377,Overvig2020Phys.Rev.B, Wang2023NC, Yan2024AOM, NLSC2025, AMSC2026}. 
These folded modes can couple to radiation, unless symmetry forbids it, as for modes that are even under $C_2$~\cite{Overvig2020Phys.Rev.B, Wang2023NC}.
Radiative $Q$ scales as $Q\propto\delta^{-2}$~\cite{Koshelev2018PRL,Wang2023NC,Overvig2020Phys.Rev.B}. The second control parameter is the lattice angle $\varphi$ [Fig.~\ref{fig:concept}(b)]. At $\varphi=90^\circ$ the lattice is rectangular and has two in-plane mirror symmetries. 
Any $\varphi\neq 90^\circ$ with $a \neq b$ removes them and makes the lattice monoclinic~\cite{Toftul2024PRL, Sinev2025,Toftul2025Nanophotonics}. Rotation $C_2$ about the $z$ axis remains for every $\delta$ and $\varphi$. The pillars sit on one side of the slab, which breaks the mirror $z\to-z$~\cite{Gorkunov2025AOM, Kumar2025ACSPhotonics, acs.nanolett.6c02271}. All the above combined, the structure therefore has no mirror symmetry and thus is chiral in a geometrical sense, although each pillar is achiral.

We calculate the eigenmodes (resonant states) of the structures with the finite-element method. It consists of TiO$_2$ pillars (refractive index 2.2, height 190~nm) on a Ta$_2$O$_5$ slab (2.1, 195~nm), between fused silica and dimethyl sulfoxide (DMSO), both 1.46, with $a=500$~nm and $b=410$~nm.
As in the fabricated samples, the pillars are truncated cones, with the base (top) diameter $20$\,nm larger (smaller) than the nominal diameter $d$ defined at mid-height.
A truncated cone has the same in-plane symmetry as a cylinder, so the taper does not change the symmetry of the structure. 
Figure~\ref{fig:knobs}(a) shows the bands near $\Gamma$ at $\delta=40$~nm and $\varphi=105^\circ$. Three states at $\Gamma$ lie on the long-wavelength side of the diffraction threshold, where only the zeroth order is open.
At $\Gamma$ the zeroth order consists of plane waves along $\pm z$. Thus, far above or below the metasurface the field of a state is uniform in the plane and transverse, $\mathbf{E}_n\simeq\ort{x}E_{n,x}+\ort{y}E_{n,y}$. Each state at $\Gamma$ is even or odd under $C_2$, that is $\hat{C}_2\mathbf{E}_n=\pm\mathbf{E}_n$ [Fig.~\ref{fig:knobs}(b)]. The rotation reverses $\ort{x}$ and $\ort{y}$, so the far field of an even state obeys $E_{n,x}=-E_{n,x}$ and $E_{n,y}=-E_{n,y}$, and it vanishes~\cite{Gorkunov2024, Kumar2025ACSPhotonics}. The state A is even, so it cannot radiate for any $\delta$ and $\varphi$, and it is a symmetry-protected BIC~\cite{PhysRevLett.127.176101, wylv-cc97}. The states \Bone{} and \Btwo{} are odd, so symmetry allows a far field, and they radiate.

\begin{figure*}[!t]
    \includegraphics[width=\linewidth]{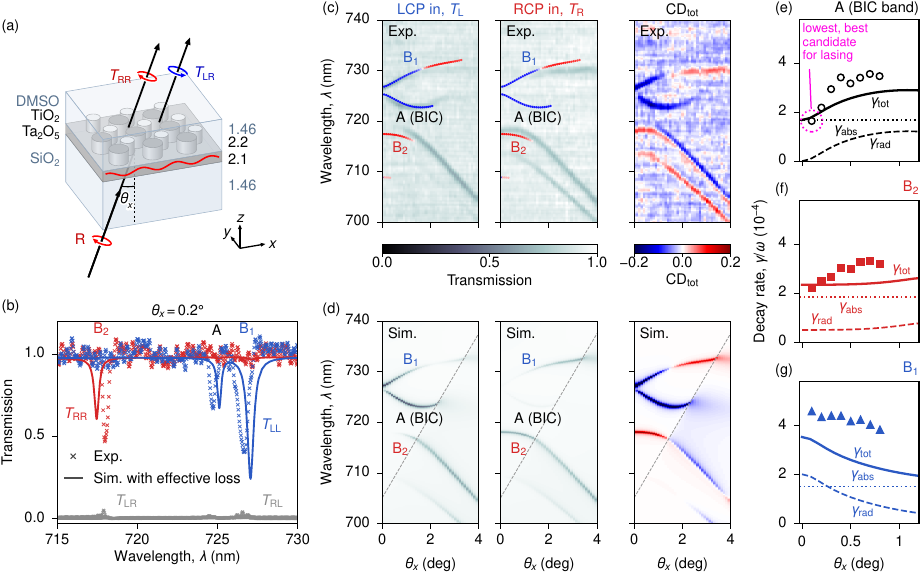}
    \caption{\label{fig:passive}
    Experimental results for linear response of Design~1 ($d=235$~nm, $\delta=40$~nm, $\varphi=105^\circ$).
    (a)~Sample and measurement scheme, with the refractive indices on the right. Light is incident from the substrate at the angle $\theta_x$ in the $xz$ plane. $T_{ij}$: transmittance into polarization $i$ for incident polarization $j$.
    (b)~Measured (crosses) and computed (lines) transmittance at $\theta_x=0.2^\circ$.
    (c)~Measured transmittance for LCP and RCP input, $T_\text{L}$ and $T_\text{R}$, and $\CDtot=(T_\text{L}-T_\text{R})/(T_\text{L}+T_\text{R})$. Dots: computed bands, colored by $\CDmode$.
    (d)~The same, computed. Dashed: onset of the first diffraction order. The color scale clips the computed $\CDtot$, which reaches 0.6.
    (e)--(g)~Decay rates of the A band, \Btwo{} and \Bone{}: computed $\grad$ (dashed), $\gabs$ (dotted) and $\gtot$ (solid); symbols: measured $\gtot/\omega = 1/(2Q)$. 
    All computed curves include the effective loss $\kappa=8.0\times10^{-4}$ (imaginary part of the refractive index).
    }
\end{figure*}

To quantify the handedness, we use the coupling parameters $m_{n\text{R}}$ and $m_{n\text{L}}$ of mode $n$ to the right- and left-circularly polarized (RCP and LCP) waves above the metasurface, and $m'_{n\text{R}}$ and $m'_{n\text{L}}$ below it~\cite{Toftul2024PRL, Gorkunov2025AOM, Kumar2025ACSPhotonics}. 
The coupling parameters give the degree of circular polarization of each radiated wave, $\stp_n = (|m_{n\text{R}}|^2 - |m_{n\text{L}}|^2)/(|m_{n\text{R}}|^2 + |m_{n\text{L}}|^2)$, and $\sbt_n$ with $m'$ in place of $m$. Both are $+1$ for RCP~\cite{PhysRevLett.124.153904}. The modal circular dichroism is then~\cite{SM,Gorkunov2024, Kumar2025ACSPhotonics, Toftul2024PRL}
\begin{equation}
    \mathrm{CD}_{\text{mode},n} = \frac{\stp_n + \sbt_n}{1 + \stp_n \sbt_n}.
    \label{eq:CD}
\end{equation}
It is $+1$ for a mode that couples only to RCP waves on both sides, and it gives the CD of the co-polarized transmission near an isolated resonance. In our design,
at $\varphi=105^\circ$, \Bone{} has $\CDmode\approx -1$ and \Btwo{} has $\CDmode\approx +1$ [Fig.~\ref{fig:knobs}(a)].
A mirror $z\to-z$ would map an RCP wave above onto an LCP wave below. Then $\sbt_n = -\stp_n$, and Eq.~\eqref{eq:CD} gives $\CDmode=0$. The one-sided pillars break this mirror.

Figure~\ref{fig:knobs}(c) varies $\delta$ at $\varphi=105^\circ$. The $Q$ factors of \Bone{} and \Btwo{} follow $\delta^{-2}$ over about three orders of magnitude, while $|\CDmode|$ stays nearly unchanged. 
To the first order in $\delta$, all four coupling parameters of a mode scale together, and their ratios do not change. 
Figure~\ref{fig:knobs}(d) varies $\varphi$ at $\delta=40$~nm. At $\varphi=90^\circ$, an in-plane mirror maps an RCP wave at $\Gamma$ onto an LCP wave, so $\stp_n=\sbt_n=0$ and $\CDmode=0$. As $\varphi$ grows, $\CDmode$ of \Bone{} and \Btwo{} approaches $-1$ and $+1$, while $Q$ changes by about a factor of two. 
The lattices with $\varphi= 90^{\circ} - \Delta \varphi$ and $\varphi= 90^{\circ} + \Delta \varphi$ are enantiomers, and have opposite $\CDmode$~\cite{Toftul2024PRL}.
Thus $\delta$ sets $Q$, and $\varphi$ sets the handedness. 

Near $\Gamma$ the A band is LCP [Fig.~\ref{fig:knobs}(a)]. For a \emph{small Bloch vector} $\mathbf{k}_B$, the modes at $\Gamma$ are the unperturbed states and $\mathbf{k}_B$ is the perturbation, as in $\mathbf{k}\cdot\mathbf{p}$ theory~\cite{Luttinger1955,Kane1957,Mei2012PRB}. The coupling of the modes $n$ and $m$ is $\propto \mathbf{k}_B\cdot\mathbf{P}_{nm}$, where $\mathbf{P}_{nm}$ is a unit-cell integral that is bilinear in their fields. A rotation by $\pi$ about a $C_2$ center reverses $\mathbf{k}_B$ and multiplies $\mathbf{P}_{nm}$ by $p_np_m$, with $p=\pm1$ the parity. It therefore multiplies the coupling by $-p_np_m$. The rotation cannot change the coupling, so it is zero unless $p_n p_m=-1$. The even state A thus mixes only with the odd B modes. The A band radiates a linearly polarized part from A itself and a B part, the sum of the B far fields weighted by coupling over detuning. In our design \Bone{} dominates, giving the A band its handedness. The A mode alone radiates linearly, so the handedness of the band comes from the odd states it couples to~\cite{Zhang2025Phys.Rev.Lett., Zhang2025arXivacp}. See more in~\cite{SM}.

We fabricate two arrays at $\varphi=105^\circ$, namely \textit{Design~1} with $d=235$~nm, $\delta=40$~nm and \textit{Design~2} with $d=255$~nm, $\delta=30$~nm. The pillars were etched into a TiO$_2$ film on a Ta$_2$O$_5$ slab on fused silica and covered with DMSO [Fig.~\ref{fig:passive}(a); see~\cite{SM} for fabrication details]. We measure the transmittance $T_{ij}$ of circularly polarized light incident from the substrate, with $i$ the output and $j$ the input polarization. Figure~\ref{fig:passive}(b) shows Design~1 at $\theta_x=0.2^\circ$. \Btwo{} appears only in $T_\text{RR}$, and A and \Bone{} appear only in $T_\text{LL}$. The cross-polarized $T_\text{LR}$ and $T_\text{RL}$ stay flat and close to zero. Figures~\ref{fig:passive}(c) and \ref{fig:passive}(d) compare the measured and computed angle-resolved transmittance. Every branch has the predicted handedness according to Eq.~\eqref{eq:CD}. \Bone{} and the A band appear for LCP input, and \Btwo{} for RCP input. The A band fades as $\theta_x\to 0$, where it becomes the BIC.

The measured dips are broader and shallower than those of the lossless structure. We describe all loss beyond the radiation of the ideal lattice, absorption and scattering by disorder alike, by one effective absorption, i.e. the indices of pillars and slab become $n+\iu\kappa$. A fit to the measured $Q$ gives $\kappa=8.0\times10^{-4}$ for Design~1 and $6.1\times10^{-4}$ for Design~2. With this loss, the computed dips come close to the measured depths [Fig.~\ref{fig:passive}(b)].

\begin{figure*}[!t]
\includegraphics[width=\linewidth]{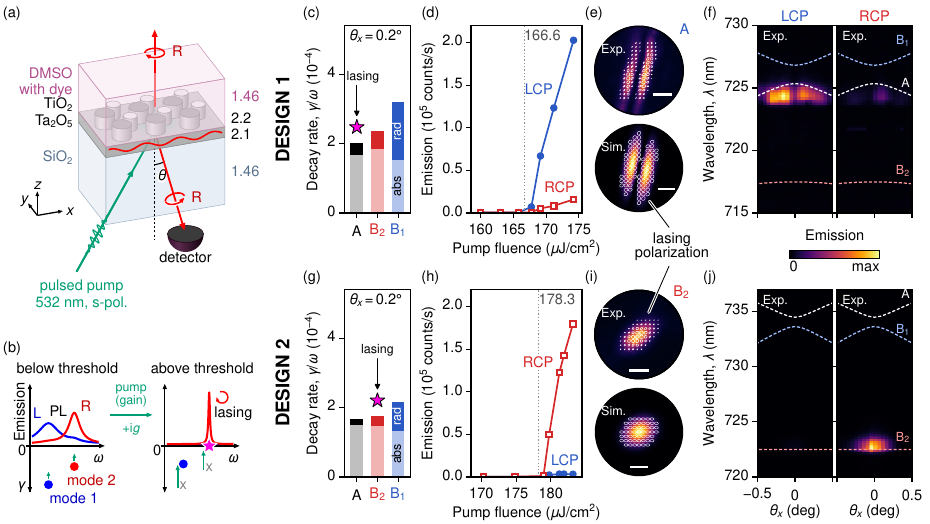}
\caption{\label{fig:lasing}
Demonstration of chiral lasing.
(a)~Measurement setup. The dye in DMSO is pumped from the substrate side by pulses at 532~nm, and the emission is collected on the substrate side at the angle $\theta$.
(b)~Lasing from resonant states. Below threshold, each mode is a pole $\Omega_n=\omega_n-\iu\gamma_n$, and the photoluminescence (PL) shows broad lines. The gain moves each pole by $\iu g_n$ [Eq.~\eqref{eq:threshold}]. The first pole to reach the real axis lases with the polarization of its mode.
(c)--(f)~Design~1. (c)~Computed decay rates at $\theta_x=0.2^\circ$, split into $\gabs$ (light) and $\grad$ (dark). Star: the lasing mode. (d)~LCP and RCP emission against pump fluence. Dotted: threshold. (e)~Far field of the laser, measured and computed from Eq.~\eqref{eq:farfield}, with the polarization ellipses of the emission. Scale bars: $0.2^\circ$. (f)~Angle-resolved emission in the LCP and RCP channels. Dashed: computed bands.
(g)--(j)~The same for Design~2.
}
\end{figure*}

Each resonant state or simply mode has the complex frequency $\Omega_n=\omega_n-\iu\gamma_n$, with $Q_n=\omega_n/(2\gamma_n)$ and the total decay rate $\gamma_n=\grad+\gabs$ ($\gtot$ in Figs.~\ref{fig:passive} and \ref{fig:lasing}). Here \( \grad \) is the decay rate associated with the radiation, while \( \gabs \) with dissipative losses. Figures~\ref{fig:passive}(e)--\ref{fig:passive}(g) decompose $\gamma_n$ into its constituent parts, assuming small losses~\cite{Muljarov2011E}.
The absorption rate $\gabs$ is similar for the three modes and nearly constant in angle, so $\grad$ sets the differences. For the A band, $\grad$ vanishes at $\Gamma$ and grows with $\theta_x$. The A band therefore has the lowest total loss near $\Gamma$, and its measured rate is lowest at $\theta_x=0.1^\circ$. Near $\Gamma$, the model follows the measured rates of A and \Btwo{}, and it underestimates that of \Bone{}. In a second set of samples ($d=255$~nm, $\delta=20$--60~nm), the measured $Q$ of \Bone{} and \Btwo{} grows as $\delta$ decreases, while the photoluminescence keeps its handedness~\cite{SM}.

For lasing, the DMSO contains the dye LDS698 (5~mM). We pump it from the substrate side with 0.52~ns pulses at 532~nm and collect the emission on the substrate side [Fig.~\ref{fig:lasing}(a)]. To first order, the gain shifts each complex frequency as~\cite{SM}
\begin{equation}
    \Omega_n \approx \omega_n - \iu\gamma_n + \iu g_n,
    \label{eq:threshold}
\end{equation}
where the gain rate $g_n$ grows with the pump and the overlap of mode $n$ with the pumped dye. Mode $n$ reaches threshold at $g_n=\gamma_n$ [Fig.~\ref{fig:lasing}(b)]. 
The mode that first reaches the threshold will become lasing.
Above threshold, the inversion is clamped, so other modes stay below their thresholds. The lasing mode is therefore the mode with the lowest $\gamma_n$. When the $\gamma_n$ of several modes are close, their overlaps with the pumped dye decide. For weak gain, the lasing field is the field of the passive mode, so the laser emits with the polarization of that mode.

In Design~1, the A band has the lowest loss near $\Gamma$ [Fig.~\ref{fig:lasing}(c)]. It lases at 724.2~nm above a threshold of $166.6\fluence$ [Fig.~\ref{fig:lasing}(d)]. The emission is LCP [Fig.~\ref{fig:lasing}(f)], with $(I_\text{R}-I_\text{L})/(I_\text{R}+I_\text{L}) = -0.88$ at the lasing peak, where $I_\text{R}$ and $I_\text{L}$ are the collected RCP and LCP intensities with an angular collection range of $\sim0.15^{\circ}$ centered at $0.2^{\circ}$.
The far field consists of two tilted lobes with a dark line through $\Gamma$ [Fig.~\ref{fig:lasing}(e)]. By $C_2$ the states at $\mathbf{k}_B$ and $-\mathbf{k}_B$ share one polarization state, so both lobes have the same handedness. The $C$ points from breaking $C_2$ have opposite handedness on the two sides of $\Gamma$~\cite{PhysRevLett.123.116104, NatureCQ2023}. A pump spot of radius $w$ makes the lasing field a packet of Bloch states with $|\mathbf{k}_B|\lesssim 1/w$. The emitted intensity in the direction $\ort{k}$ set by $\mathbf{k}_B$ is then~\cite{SM}
\begin{equation}
    I(\ort{k}) \propto |\psi_n(\mathbf{k}_B)|^2 \left( |m'_{n\text{R}}(\mathbf{k}_B)|^2 + |m'_{n\text{L}}(\mathbf{k}_B)|^2 \right),
    \label{eq:farfield}
\end{equation}
where $\psi_n(\mathbf{k}_B)$ is the amplitude of the Bloch state $\mathbf{k}_B$ in the lasing field.
Near $\Gamma$ the band is quadratic $\omega_n(\mathbf{k}_B)\approx\omega_n(\mathbf{0})+D_1K_1^2+D_2K_2^2$, with $K_{1,2}$ the components of $\mathbf{k}_B$ along its principal axes. 
All components oscillate at the lasing frequency $\omega_L \approx \omega_n(\mathbf{0})$, so the component $\mathbf{k}_B$ is detuned from its own frequency by $D_1 K_1^2 + D_2 K_2^2$.
For the gain profile $g_n \eu^{-\rho^2/w^2}$, with $\rho$ the distance from the center of the pump spot, the packet extends in $\mathbf{k}$ until this detuning, $|D_i| K_i^2$, equals the gain lost at the edge of the mode, $g_n/(w K_i)^2$.
With $g_n=\gamma_n$ at threshold, the packet is a Gaussian with an effective radius $w_i$ along each axis,
$|\psi_n(\mathbf{k}_{B})|^{2} \propto \exp \left( - K_1^2 w_1^2 - K_2^2 w_2^2 \right)$ with $w_i^{2}=\sqrt{w^2|D_i|/(2\gamma_n)}$~\cite{SM}.
The sum $|m'_{n\text{R}}|^2+|m'_{n\text{L}}|^2$ is the rate at which the mode loses energy into the substrate~\cite{Fan2003JOSAAJ, Toftul2024PRL, Gorkunov2025AOM}. For the A band it vanishes at $\Gamma$, which gives the dark line. The complex ratio $m'_{n\text{R}}/m'_{n\text{L}}$ fixes the polarization ellipse in each direction, and its modulus gives $\sbt_n(\mathbf{k}_B)$. Equation~\eqref{eq:farfield} reproduces the two lobes, their tilt, and their handedness. We take $w=160$~\textmu m, which gives the measured lobe separation of $0.17^\circ$.
In experiment, the pump spot diameter is $124$~\textmu m. Due to the nonlocal nature of the metasurface modes, the real excitation size can be larger than that, as the size of the fabricated array is roughly $400\times400$~\textmu m.
Averaged over the far field, the model gives $\sbt_n=-0.97$, against the measured $-0.88$.

Design~2 has a smaller $\delta$, which raises the radiative $Q$ of \Btwo{} as $\delta^{-2}$. Its three modes now have close total losses [Fig.~\ref{fig:lasing}(g)]. \Btwo{} lases at 722.7~nm above a threshold of $178.3\fluence$ [Fig.~\ref{fig:lasing}(h)]. The calculated loss of A is slightly lower than that of \Btwo{}. At such small differences, the overlap of each mode with the pumped dye follows the pump field inside the structure. The laser radiation is RCP, with $(I_\text{R}-I_\text{L})/(I_\text{R}+I_\text{L}) = +0.96$ at the lasing peak, and Eq.~\eqref{eq:farfield} gives $\sbt_n=+0.88$. The laser emits a single lobe at $\Gamma$, because \Btwo{} radiates there [see Figs.~\ref{fig:lasing}(i) and \ref{fig:lasing}(j)]. Michelson interferometry confirms the coherence of both lasers~\cite{SM}. The two designs thus lase with the opposite handedness, and each laser takes its handedness from its corresponding mode.

In conclusion, we have demonstrated that two parameters of a monoclinic lattice composed of achiral pillars placed on a slab can control almost separately two properties of a chiral resonance. The dimerization $\delta$ sets $Q$ through band folding, and the lattice angle $\varphi$ sets the handedness through the mirror symmetries. 
The remaining $C_2$ symmetry protects BIC at the $\Gamma$ point, whose band becomes chiral by mixing with the bright modes near $\Gamma$.
Thus, with only single monoclinic metasurface, we have achieved LCP lasing on the BIC band and RCP lasing on a bright mode. The separation relies on the symmetries of the lattice and on bright modes coupled to the BIC band, so our concept applies to other materials and wavelengths, and it can be employed for nonlinear and quantum chiral sources.

\vspace{0.5\baselineskip}
\begin{acknowledgments}
X.A. acknowledges support from the National Natural Science Foundation of China (Grant Nos. 12374310). Y.C. acknowledges support from the National Natural Science Foundation of China (Grant Nos. 12534014, 12192254, 92250304, W2441005) and the National Key Research and Development Program of China (Grant No. 2022YFA1404800).
\end{acknowledgments}

\paragraph{Data availability.}
The data that support the findings of this article are openly available~\cite{li2026chiraldata}.

\paragraph{AI use statement.}
The authors used Claude Code with the Claude Opus 5.5 to improve text, to write the simulation and post-processing scripts. The authors directed this work, checked every equation and statement, and take full responsibility for the content.

\bibliography{references}

\end{document}